\documentclass[conference]{IEEEtran}
\ifCLASSINFOpdf
   \usepackage[pdftex]{graphicx}
\else
\fi
\usepackage{amsmath}
\usepackage{algorithmic}

\usepackage{booktabs}
\usepackage{threeparttable}

\usepackage[hide]{todo}

\begin{document}
%
\title{A Faster DiSH: Hardware Implementation of a Discrete Cell Signaling Network Simulator}

\author{\IEEEauthorblockN{Kevin Gilboy, Khaled Sayed, Niteesh Sundaram, Kara Bocan, and Natasa Miskov-Zivanov}
\IEEEauthorblockA{
University of Pittsburgh\\
Pittsburgh, PA, USA, 15213\\
Email:nmzivanov@pitt.edu}
}


%


\maketitle

\begin{abstract}
Development of fast methods to conduct \textit{in silico} experiments using computational models of cellular signaling is a promising approach toward advances in personalized medicine. However, software-based cellular network simulation has runtimes plagued by wasted CPU cycles and unnecessary processes. Hardware-based simulation affords substantial speedup, but prior attempts at hardware-based biological simulation have been limited in scope and have suffered from inaccuracies due to poor random number generation. In this work, we propose several hardware-based simulation schemes utilizing novel random update index generation techniques for step-based and round-based stochastic simulations of cellular networks. Our results show improved runtimes while maintaining simulation accuracy compared to software implementations.
\end{abstract}


%
\IEEEpeerreviewmaketitle

\section{Introduction}
Biological systems are stochastic in nature; biochemical reactions have a certain probability of occurring, even when the concentrations of reactants are known, and the environmental conditions are controlled. This is in direct contrast to a deterministic system, in which a given set of inputs will always produce the same set of outputs. Computational models can mimic the randomness of biological systems to gain insight into the nature of particular biological mechanisms, such as the role a specific ligand plays in a signaling network, or the substance concentration needed to initiate a reaction. In addition, computational models have been shown to be invaluable as researchers can run a large number of \textit{in silico} experiments that would be inefficient or impractical if attempted \textit{in vivo} or \textit{in vitro}. Such models allow for predicting wet-lab outcomes and for shedding insights into the nature of biological systems. 

As computational models are versatile and can be constantly updated via input from the user, they have the potential to be used in clinical applications such as  personalized medicine. A model of a physiological system could be created and then modified in real time to more accurately reflect the unique characteristics of a specific person. The model could then be used to predict patient outcomes and inform treatments. 

Although many such personalized medicine software (SW) models exist, they are slower and less efficient than hardware (HW) approaches \cite{salwinski2004_insilico, yoshimi2007_fpga, miskov-zivanov2011_regulatory, miskov-zivanov2011_emulation}. Implementation in FPGAs allows running concurrent models at a speed superior to current SW models, while still maintaining accuracy and reliability. In this paper, we describe HW emulation of a discrete model of T cell differentiation in SystemVerilog, and compare the accuracy and runtime with SW simulation for different simulation schemes and scenarios. These simulation schemes and scenarios represent stochasticity in cellular signaling networks and different component conditions and changes that occur in these networks. Selecting elements for update in such simulations is a non-trivial task, therefore we also discuss and implement novel random update index generation techniques for stochastic simulation schemes.


%


\section{Methods}


A T cell differentiation model 
was used in this work to compare HW and SW simulation. The model is described in \cite{miskov-zivanov2013_duration}, where it was analyzed using the SW simulator described in \cite{albert2008_boolean}. In this work, we compared HW and SW simulation of this model using simulation schemes described in \cite{sayed2017_dish}, where an executable model consists of elements or element groups that are updated simultaneously within the group. Each element has an associated update rule, and elements/groups have a certain probability of their rules being executed in a given simulation step or round. Here we implemented: the \textit{simultaneous} (SMLN) scheme; \textit{random-order sequential} (RSQ) scheme, both \textit{round-based} (RB-RSQ) and \textit{step-based} (SB-RSQ); and \textit{grouped random-order sequential} (RSQ-g) scheme, both \textit{round-based} (RB-RSQ-g) and \textit{step-based} (SB-RSQ-g). 
The HW framework and simulation schemes were implemented in SystemVerilog and simulated using ModelSim. 
For the SW comparison, we used the DiSH (Discrete, Stochastic, Heterogeneous) SW simulator, previously described in \cite{sayed2017_dish}.  



\begin{figure}[!t]
\centering
\includegraphics[width=3.1in]{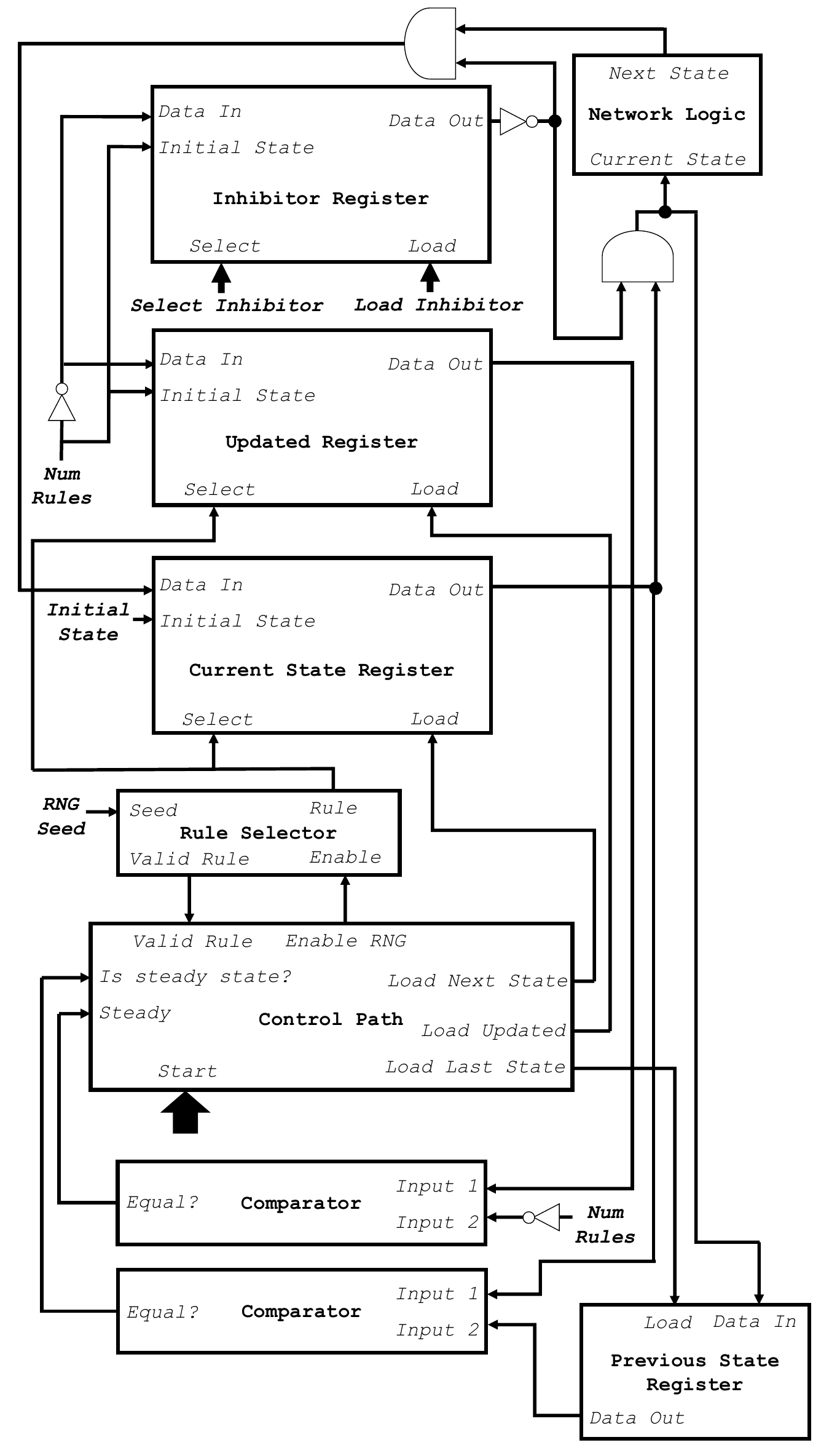}

\caption{Hardware framework for simulating cellular networks with different simulation schemes and random update index generation.}
\label{fig:simulator_hw}
\end{figure}

\subsection{HW Framework}

A schematic of the HW simulator is shown in Figure \ref{fig:simulator_hw}. The simulation schemes are implemented and selected within the \texttt{\textbf{Rule Selector}} block, and each simulation is run for a specified number of steps or rounds. In addition, the model can be initialized with different element values at the beginning of each simulation. 

A \texttt{\em Start} signal to the \texttt{\textbf{Control Path}} module prompts the \texttt{\textbf{Rule Selector}} to randomly generate an index via the \texttt{\em Enable} pin. If the \texttt{\em Valid Rule} pin indicates a valid index, \texttt{\textbf{Control Path}} asserts a high \texttt{\em Load} signal to the \texttt{\textbf{Inhibitor}}, \texttt{\textbf{Updated}}, and \texttt{\textbf{Current State Registers}}, allowing their values to be modified on the next clock cycle. The index is used as a \texttt{\em Select} input for the \texttt{\textbf{Current State Register}}, which keeps track of the state of each element in the model. The output (\texttt{\em Data Out}) of the \texttt{\textbf{Current State Register}} is then bitmasked with output (\texttt{\em Data Out}) from the \texttt{\textbf{Inhibitor Register}}, where the inhibitor is chosen by the initial conditions (\texttt{\textbf{\em Select Inhibitor}}). The result of the combined current state and inhibition is stored in the \texttt{\textbf{Previous State Register}} and passed into the \texttt{\textbf{Network Logic}} module, which contains the update rules of the model. The current state is passed on the \texttt{\em Current State} pin, and the \texttt{\textbf{Network Logic}} module outputs the next state of the system on the \texttt{\em Next State} pin. The result is stored in the \texttt{\textbf{Current State Register}}. The circuit is ``steady'' once all rules have been run, as tracked by the \texttt{\textbf{Updated Register}}. The upper \texttt{\textbf{Comparator}} module compares the updated elements/groups with the complete list of elements/groups, asserting the \texttt{\em Steady} pin on the \texttt{\textbf{Control Path}} if they are equal. The circuit is said to be in ``steady state'' if the current state is equal to the previous state. The lower \texttt{\textbf{Comparator}} module compares the output of the \texttt{\textbf{Current State Register}} to that of the \texttt{\textbf{Previous State Register}} and asserts the \texttt{\em Is steady state?} pin on the \texttt{\textbf{Control Path}} if they are equal. 

\subsection{RNG Algorithms}

While the SMLN scheme is deterministic, all other schemes require random number generation (RNG) to generate a random rule index that selects elements/groups and executes their update rules in each simulation round/step. We describe here two novel HW-based algorithms for RNG that we created and implemented within the \texttt{\textbf{Rule Selector}} block in Figure \ref{fig:simulator_hw}. 

\subsubsection{Round-based RNG}

\begin{figure*}[!t]
\centering
\includegraphics[width=0.85\textwidth]{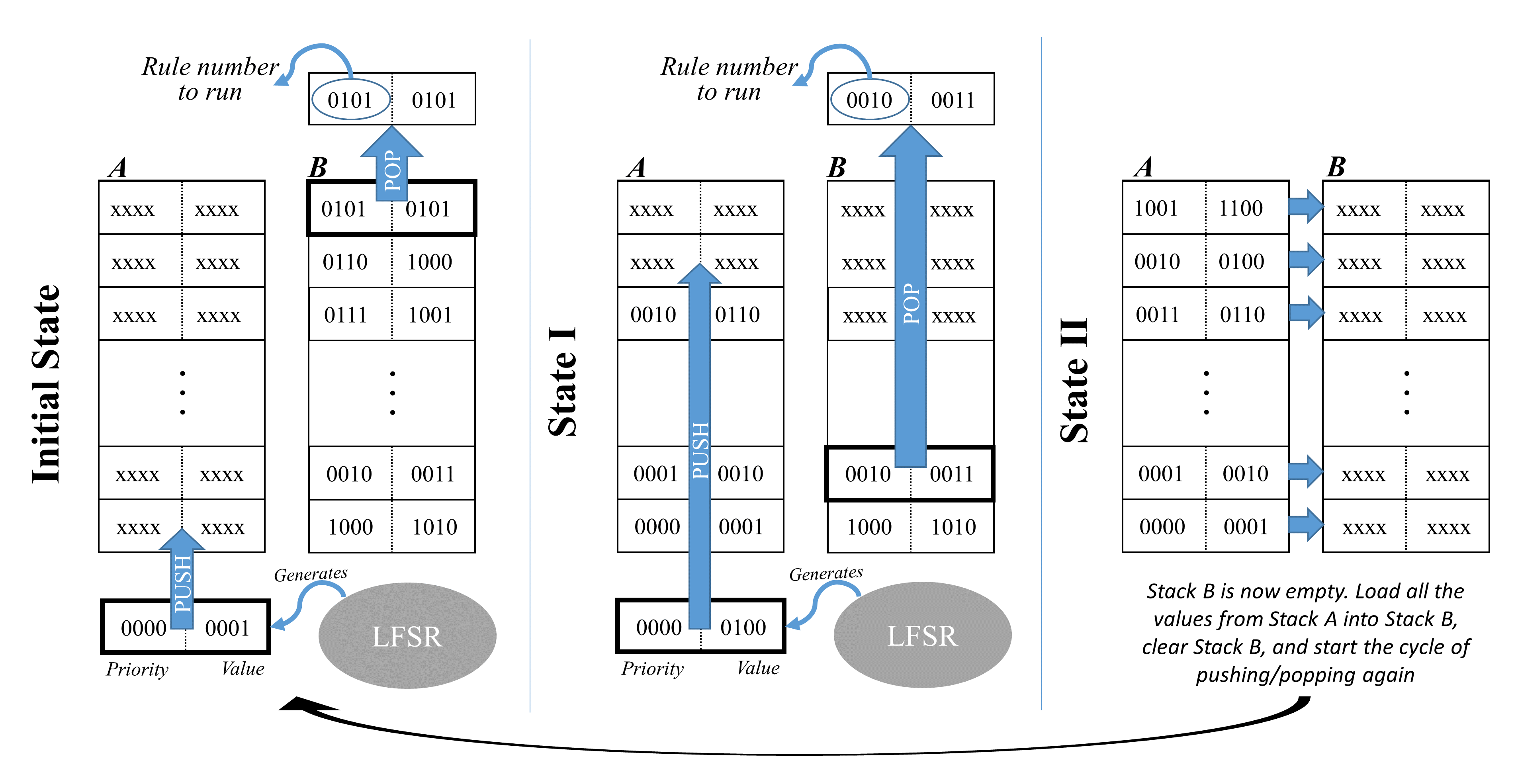}
\caption{The RB-RSQ rule selection scheme implementation for a sample 16-rule system, utilizing a linear feedback shift register (LFSR). A 16-element/group system as diagrammed here would be $2\log_2 16$ = 8 bits wide and 16 registers tall. Each register in the array could be thought of as being logically divided in half, with the least significant bits corresponding to a Value, and the most significant bits indicating the Priority. 
}
\label{fig:rng}
\end{figure*}

In the round-based simulation schemes, each element/group is updated at least once in a given round; therefore each round consists of a number of steps equal to the number of elements/groups in the model. Our round-based RNG algorithm allows linear runtime with respect to the number of elements/groups in the model by eliminating the possibility of duplicate or invalid rule index generation. The operation is illustrated in Figure \ref{fig:rng}. Two parallel register arrays, A and B, are utilized as stacks for storing items consisting of a rule index (``Priority'') and randomly generated number (``Value"). Stack A always starts empty and grows upward while Stack B starts out full and recedes downward. With each step, a new item is pushed up Stack A. 
If the new item's Value is greater than an existing Value, the Priority of the new item is increased by one. Otherwise, the Priority of the existing item is increased by one. If there are no other Values on the stack, the rule takes a default priority of 0. Simultaneously to the generation of Stack A, an item is popped off of B, and its Priority determines which element/group is updated in the current step. When a round is completed, Stack B is empty and Stack A is full. The items from A are then loaded into B, Stack A is flushed, and the parallel generation and popping of items repeats. At the start of the simulation, where both stacks are empty, items must be generated to fill Stack B before a round can begin. After the initial generation of Stack B, the coincident generation and popping of items allows linear runtime without duplicates or invalid rules.

%



\subsubsection{Step-based RNG}

A RNG algorithm was also developed for the step-based scheme, where the same element/group can potentially be updated multiple times before other elements/groups are updated. Directly using the RNG number as the rule index is highly inefficient for a model that has a number of elements/groups not equal to a power of two: a model with 37 elements would need 6 bits to represent each of the potential element updates, leading to $2^6 - 37 = 27$ RNG outputs (42\%) counting as a costly ``miss''. Here, we determine the rule index ($I$) using Equation \ref{eq:modular_index}, where $X$ is the RNG number and $E$ is the number of elements/groups. We also use 10 RNG bits ($n$ = 10) to approach uniform probability of generating any index, calculated by setting a threshold of $(2^n\mod{E})/2^n \textless 3\%$. 
%

\begin{equation}
\label{eq:modular_index}
I = X\mod{E}
\end{equation}

\subsection{HW and SW Comparison}

We simulated eight scenarios representing different input value configurations, and compared the SW and HW simulators by studying the responses of FOXP3 and IL2 (markers of regulatory or helper T cell phenotypes, respectively). Table \ref{tab:initial} shows the initial values for input signals that were varied for the eight scenarios: antigen dose affecting T cell receptor (TCR) signal strength, transforming growth factor beta ligand (TGF), and the inhibitor of protein kinase B (AKT\_off). 
The percentages listed for the Toggle scenarios in Table \ref{tab:initial} represent the percentage of simulation steps/rounds that were completed before the protein's value was toggled. Of the variables not listed in Table \ref{tab:initial}, CD28, PTEN, TSC, CD122, and CD132 were initialized to 1, and all the other variables were initialized to 0, to mimic the na\"{i}ve T cell phenotype at the beginning of each simulation. We ran all round-based simulations for 30 rounds, and all step-based and SMLN simulations for 2000 steps \cite{miskov-zivanov2013_duration}. We ran each simulation scheme 200 times from the initial to steady state and calculated average trajectories, according to methodology from \cite{sayed2016}.

\begin{table}[!t]
\centering
\begin{threeparttable}
\renewcommand{\arraystretch}{1.3}
\caption{Simulation scenarios and initial values}
\label{tab:initial}
\begin{tabular}{c c c c c c}
\toprule
Scenario 	& 	TCR\_high	&	TCR\_low	&	TGFbeta	&	AKT\_off	&	Toggle	\\
\midrule
1		&	1		&	0		&	0		&	0		&	-		\\
2		&	0		&	1		& 	0		&	0		&	-		\\
3		&	1		&	0		&	1		&	0		&	-		\\
4		&	0		&	1		&	1		&	0		&	-		\\
5		&	1		&	0		&	0		&	1		&	-		\\
6		&	1*		&	0		&	0		&	0		&	20.00\%	\\
7		&	1*		&	0		&	0		&	0		&	26.67\%	\\
8		&	1*		&	0		&	0		&	0		&	33.33\%	\\
\bottomrule
\end{tabular}
\begin{tablenotes}
\item* Initial value before toggle
\end{tablenotes}
\end{threeparttable}
\end{table}

\section{Results}

\begin{figure}[!t]
\centering
\includegraphics[width=3.5in]{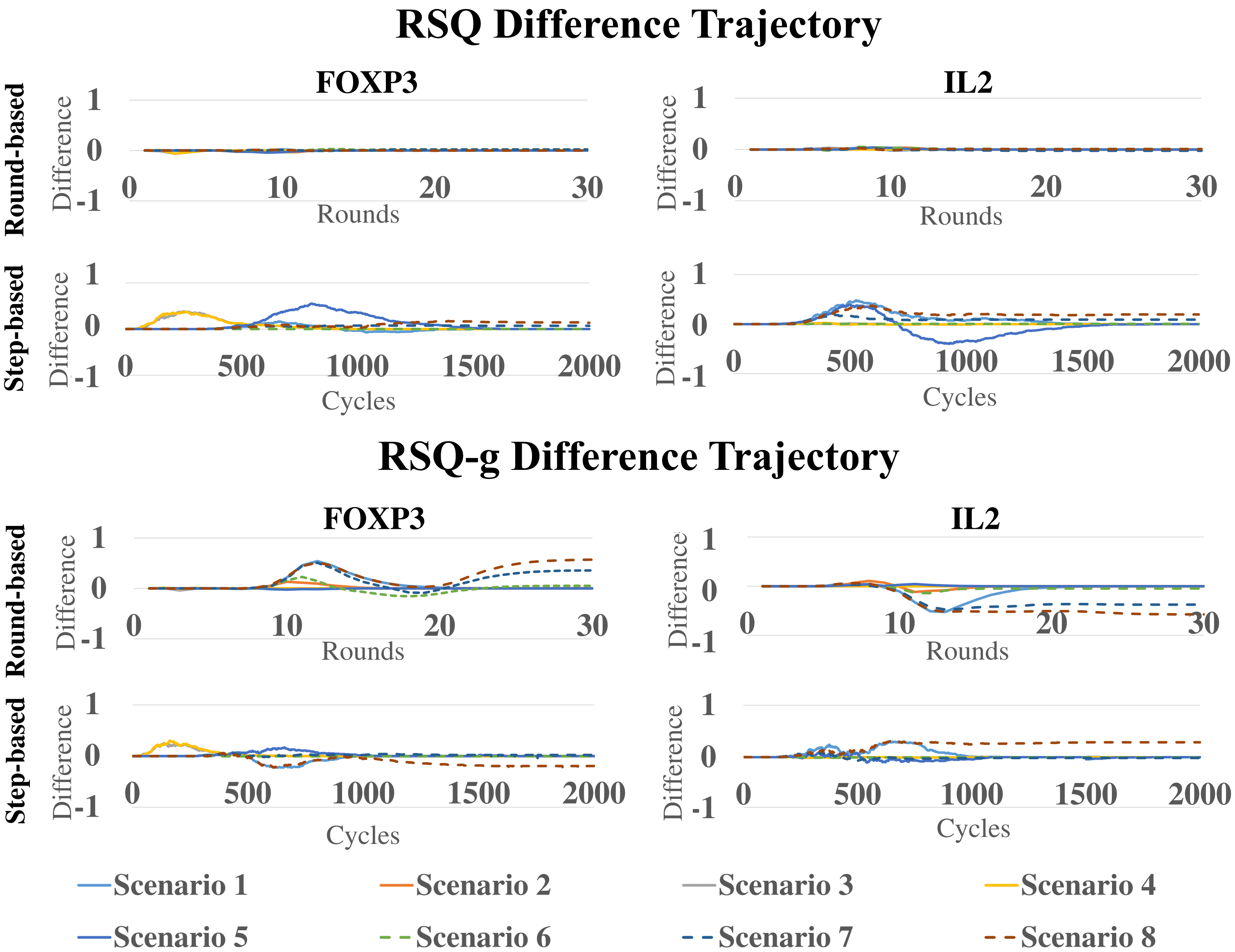}
\caption{Difference between HW and SW simulator output for FOXP3 and IL2 with the RSQ scheme (top), and the RSQ-g scheme (bottom).}
\label{fig:comparison}
\end{figure}
\begin{figure}[!t]
\centering
\includegraphics[width=3.5in]{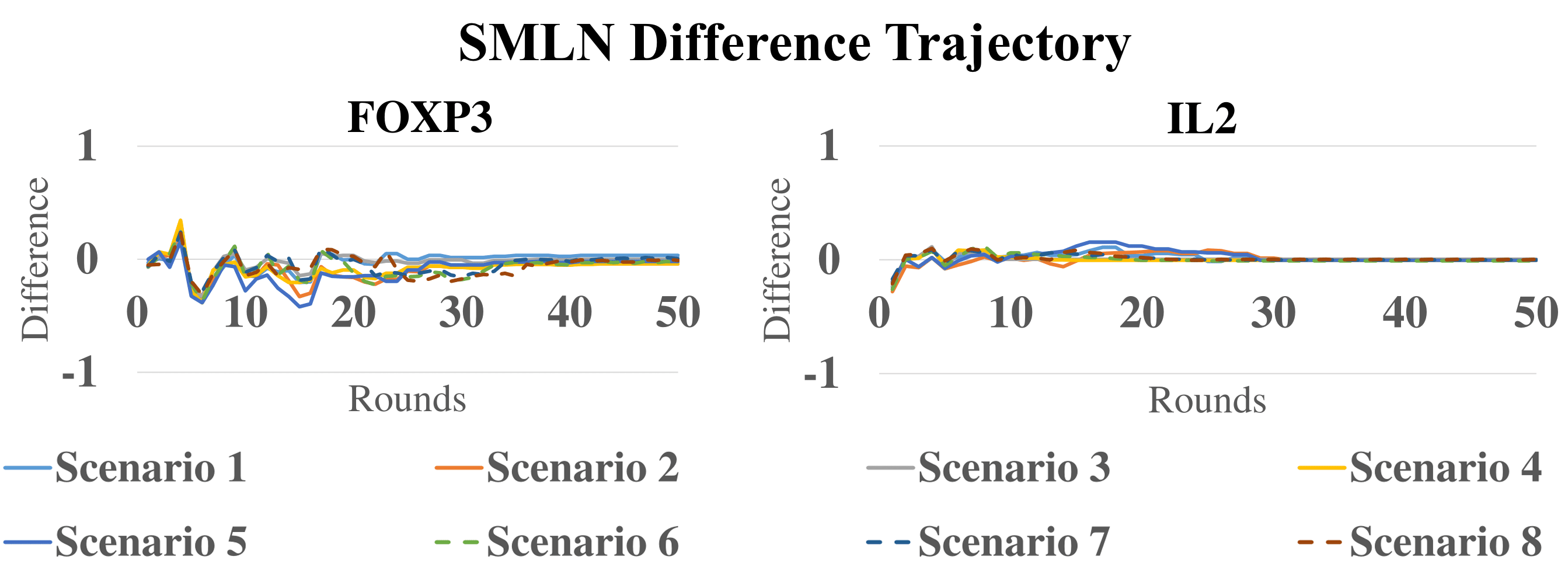}
\caption{Difference between HW and SW simulator output for FOXP3 and IL2 with the SMLN scheme.}
\label{fig:comparison_sim}
\end{figure}

Figures \ref{fig:comparison} and \ref{fig:comparison_sim} show the difference between average trajectories obtained with the HW simulator and the SW simulator for FOXP3 and IL2 and each simulation scheme. Average root-mean-square error (RMSE) across all scenarios for each simulation scheme is shown in Table \ref{tab:rmse}.

%

\begin{table}[!t]
\renewcommand{\arraystretch}{1.3}
\caption{RMSE averaged across all simulation scenarios}
\label{tab:rmse}
\centering
\begin{tabular}{r r r r r r}
\toprule
		&	\multicolumn{5}{c}{Scheme} 	\\
\% Error		&	RB-RSQ		&	SB-RSQ	&	RB-RSQ-g		& 	SB-RSQ-g	&	SMLN 		\\
\midrule
FOXP3	& 	1.13		&	11.18	&	10.86		&	6.22		&	14.99	\\
IL2		&	0.99		&	10.79	&	13.00		&	5.41		&	6.23		\\
\bottomrule
\end{tabular}
\end{table}

Runtime comparisons are shown in Table \ref{tab:runtimes} for each simulation scheme. For HW runtimes, the number of clock cycles used as reported by ModelSim was divided by 50 MHz, a common FPGA base clock frequency. This was compared to SW runtime as performed on a 2015 Apple MacBook Pro (3.1 GHz Intel i7 processor). The HW implementation had a median speedup of 54.4X for round-based simulations, and 426.7X for step-based simulations. The greatest speedup was provided for the SMLN scheme, followed by RSQ-g and then RSQ.

\section{Discussion}

Our results show that all five simulation schemes in HW produced results comparable to the SW simulator, with all RMSEs under 15\%. 
The round-based implementation had the smallest RMSE, but with the smallest runtime speedup, making it a Las Vegas type algorithm. Conversely, the SMLN scheme showed the greatest runtime speedup but greatest RMSE, making it a Monte Carlo type algorithm. The grouped step-based implementation had the second smallest RMSE and the fastest runtime speedup for all the RSQ methodologies. The step-based HW implementation was expected to provide greater speedup than the round-based implementation because the round-based RNG relies on pushing numbers onto a stack at linear runtime. Similarly, the SMLN schemes show much greater speedup than the other schemes, because every element's value is updated in each step without the need for RNG. However, the RSQ schemes are more desirable for modeling stochasticity in biological networks \cite{salwinski2004_insilico, yoshimi2007_fpga, miskov-zivanov2011_regulatory, miskov-zivanov2011_emulation}. The speedups for RSQ simulation schemes were similar to previous stochastic simulation work by Yoshimi et al. \cite{yoshimi2007_fpga}, and HW runtimes were similar to those reported in \cite{miskov-zivanov2011_emulation}, despite differences in simulator architectures. 

All non-Toggle HW simulations approached zero difference from the SW simulator at steady state. For the Toggle scenarios, a large deviation occurred subsequent to the toggle of the antigen dose, and the steady state values were different between the SW and HW simulators, particularly for the grouped schemes. This can be likened to a logic circuit whose inputs change before the output has stabilized. 

\begin{table}[!t]
\renewcommand{\arraystretch}{1.3}
\caption{Runtime comparison for each simulation scheme.}
\label{tab:runtimes}
\centering
\begin{tabular}{r r r r r}
\toprule
		&				&	\multicolumn{2}{c}{Runtime [ms]}		&			\\
Scheme	& 	Steps/Rounds		& 	SW 				&	HW 			&	Speedup 	\\
\midrule
\textit{Round-Based Schemes} &  & & & \\
RB-RSQ		& 	1929500		&	1062.5 			&	38.6			&	27.5X	\\
RB-RSQ-g 	&	1216500		&	1253.6			&	24.3			&	51.5X	\\
SMLN		&	11300		&	645.8			&	0.2			&	2857.5X	\\
RB-RSQ Toggle &	1929500		&	1027.6			&	38.6			&	26.6X	\\
RB-RSQ-g Toggle &1216500		&	1394.7			&	24.3			&	57.3X	\\
SMLN Toggle &	11300		&	688.9			&	0.2 			&	3048.3X	\\
\textit{Step-Based Schemes} &  & & & \\
SB-RSQ		&	401300		&	3225.1 			&	8.0				&	401.8X	\\
SB-RSQ-g	&	303537		&	2745.0			&	6.1				&	452.2X	\\
SB-RSQ Toggle &	401300		&	3036.9			&	8.0				&	378.3X	\\
SB-RSQ-g Toggle &303537		&	2741.7			&	6.1				&	451.6X	\\
\bottomrule
\end{tabular}
\end{table}

The most likely source of error in the RSQ methodologies lies in comparing the randomness of the HW RNG and the randomized SW update rule selection. This is evidenced by the greater differences in transient behavior compared to steady state. Comparing the HW and SW simulators with the same sequence of element/group updates could provide more similar transient results, and could be investigated in future work.

The SMLN scheme implementations showed differences at the start of the simulation, but reached zero difference at steady state. 
In the SMLN scheme, each element not governed by a forced initial condition was given a random initial state, leading to $2^{52}$ possible initial states. Since only 200 of these initial states were tested, and the 200 states used in the HW model most likely differed from the 200 states used in the SW model, deviations are expected especially at the start of the simulation due to different initial conditions. In a simulation where the HW and SW models utilize the same initial states, the HW and SW simulation results are the same.


\section{Conclusion}

In this work, we designed and emulated five HW-based simulation schemes and two random update index generation techniques that were accurate in predicting phenotype decisions in T cells and afforded orders of magnitude runtime speedup compared to SW simulation. For the T cell model in this work, the round-based implementation was most accurate to SW but provided the smallest runtime speedup, while the SMLN implementation provided the fastest result although with lower accuracy to SW. Future work includes parallelization, as well as improving the random update index generation to increase the accuracy without the cost of runtime.

\section*{Acknowledgment}

This work is supported in part by DARPA Big Mechanism award W911NF-14-1-0422.



\bibliographystyle{IEEEtran}
\bibliography{references}
%
%
%

\end{document}